\documentstyle[12pt]{article}
\makeatletter
\newcommand{\singlespacing}{\let\CS=\@currsize\renewcommand{\baselinestretch}
{1.0}\tiny\CS}
\newcommand{\doublespacing}{\let\CS=\@currsize\renewcommand{\baselinestretch}
{1.5}\tiny\CS}

\newcommand {\beq} {\begin{equation}}
\newcommand {\al} {\alpha}
\newcommand {\eeq} {\end{equation}}
\newcommand {\bt} {\beta}
\newcommand {\om} {\omega}
\newcommand {\dl} {\delta}
\newcommand {\Dl} {\Delta}
\date{}
\begin{document}

\thispagestyle{empty}\setcounter{page}{1}
\vskip10pt
\centerline{\bf SHIFT OF SPECTRAL LINES DUE TO
 DYNAMIC MULTIPLE SCATTERING}
\centerline{ \bf AND SCREENING EFFECT : IMPLICATIONS FOR DISCORDANT REDSHIFTS}

\vskip20pt
\centerline{\footnotesize Sisir Roy$^{1,2}$, \& 
Menas Kafatos$^1$  }
\centerline{\footnotesize \it Center for Earth Observing and Space Research}
\centerline{\footnotesize \it Institute for Computational Sciences and 
Informatics and}
\centerline{\footnotesize \it Department of Physics, George Mason University} 
 \baselineskip=10pt
\centerline{\footnotesize \it Fairfax,  VA  22030  USA}
\vskip10pt
\centerline{\footnotesize Suman Datta$^2$ }
\centerline{\footnotesize \it Physics and Applied Mathematics Unit}
\centerline{\footnotesize \it Indian Statistical Institute, Calcutta, INDIA}
\vskip5pt
\centerline{\footnotesize $^{1,2}$ e.mail: sroy2@osf1.gmu.edu  }
\centerline{\footnotesize $^1$ e.mail : mkafatos@compton.gmu.edu}
\centerline{\footnotesize $^2$ email : res9428@isical.ac.in }

\vskip20pt

\doublespacing
\abstract{\noindent{\small{  The frequency shift of spectral lines from
astronomical objects
is most often explained by the Doppler Effect arising in  relative motion and
 the 
 broadening
of a particular line  is supposed to depend on the absolute temperature ,
 pressure 
or the different line of sight velocities.
The Wolf effect on the
other hand deals with correlation induced spectral changes and explains
both the broadening and shift of the spectral lines. In this framework
a sufficient condition for redshift has been derived and when applicable
the shift is shown to be larger than broadening. Under this condition of larger
shift than broadening we find a critical source frequency below which
no spectrum is analyzable for a particular medium. This gives rise to
new type of screening effect which may play a  significant role both at laboratory
scale as well as in the astronomical domain. We apply a simple interpretation
of the discordant redshifts in galaxy-quasar associations.
\vskip10pt
\noindent

\newpage

\section{{\bf Introduction}}

In studying the motion of astronomical objects, astrophysicists utilize the 
study of frequency  shift of spectral lines.
A redshift of spectral lines is the phenomenon of displacement of the lines
towards longer wavelengths. Although for a lot of astronomical objects
such as stars, redshifts and blueshifts are easily understood in terms of
relative motion, the origin of redshift  of quasars has been one of the
most controversial topics for the last few decades. Interpretation of redshift
as Doppler shift has been broadly accepted, although this could not solve the 
observed anomalies in the
quasar redshifts and some related problems(Wolf,1998 ).

Recent results in Statistical Optics ( Wolf,1986 ) has made it clear that the shift of
the frequency of the spectral lines can be explained without considering
the relative motion of the observer with respect to the source.
Following this idea, dynamic multiple scattering theory(Datta et al. 1998a,b )  has been
developed to account for the shift  as well as the
broadening of the  spectral line. It
is shown that when light passes through a turbulent(or inhomogeneous ) medium,
due to multiple scattering effects the shift and the width  of the line can be calculated.
Here, a sufficient condition for redshift has been derived and when applicable
 the shift is shown to be larger than broadening. The width of the
spectral line can be calculated after multiple scatterings and a relation
can be derived between the width and the shift $z$ which applies to
any value of z (Roy et al.1999). 

Using the above condition it is shown that there exists a lower bound of
the source frequency below which no spectra is analyzable in the sense that
the broadening will be  larger than the shift of the spectral line and the
identity of the line becomes confused. 
This critical source frequency depends very much on  the actual  nature of the medium
through which the light is propagating. We call this  the screen effect
due to the medium. 
This new screen effect may play an important role in astronomical domain.
This can also be verified   in  laboratory experiments.

We first  discuss briefly the theory of multiple scattering
in section 2. In section 3 we shall introduce the critical source frequency and its
implications while in section 4 we shall discuss  a special type of
screening induced by critical source frequency. Finally, a general discussion
will be made on the possible applications of this type of screening.

\vskip10pt
\noindent
\section{{\bf Dynamic Multiple Scattering Theory}}

First, we briefly state the main results of Wolf's scattering mechanism. Let us
consider a polychromatic electromagnetic field of light of central frequency $\omega_0$ 
and 
width $\delta_0$, incident on the scatterer. The incident spectrum is assumed to be of the
form
\begin{equation}\displaystyle{
S^{(i)}(\omega)= A_0e^{\left[-\frac{1}{2\delta_0^2}(\omega-\omega_0)^2\right]}
}
\end{equation}
The spectrum of the scattered field is given by (Wolf 
and James,1996 )
\begin{equation}\displaystyle{
S^{(\infty)}(r \vec{u'} ,\omega')=A\omega'^4\int_{-\infty}^{\infty}
K(\omega,\omega',\vec{u}, \vec{u'})S^{(i)}(\omega)d\omega}
\end{equation}

which is valid within the first order Born approximation (Born and Wolf ,1997 ). Here $K(\omega, \omega')$ is the so called scattering kernel and it
plays the most important role in this mechanism. Instead of studying $\cal
K(\omega,\omega')$ in detail, we consider a particular case for the
correlation function $G(\vec{R},T;\omega) $ of the generalized dielectric
susceptibility $\eta(\vec{r},t;\omega)$ of the medium which is characterized by an
anisotropic Gaussian function
\begin{equation}\displaystyle{
\begin{array}{lcl}
G(\vec{R},T;\omega) & = &
<\eta^*(\vec{r}+\vec{R},t+T;\omega)\eta(\vec{r},t;\omega)>\\ \\
& = &G_0exp\left[-\frac{1}{2}\left(\frac{X^2}{\sigma_x^2}
+\frac{Y^2}{\sigma_y^2}
+\frac{Z^2}{\sigma_z^2}+\frac{c^2T^2}{\sigma_\tau^2}\right)\right]
\end{array}
}
\end{equation}
\vskip5pt
\noindent
Here $~G_0~$ is a positive constant, $~\vec{R}~=~(X,~Y,~Z)~$, and
$~\sigma_x~,~\sigma_y~,~\sigma_z~,~\sigma_\tau~$ are correlation 
lengths. The anisotropy is indicated by the unequal
correlation lengths in different spatial as well as temporal directions. $\cal
K(\omega,\omega')$ can be obtained from the four dimensional Fourier Transform
of the correlation function $G(\vec{R},T;\omega) $.  In this case $\cal
K(\omega,\omega')$ can be shown to be of the form
\begin{equation}\displaystyle{
{\cal{K}} (\omega,\omega')=B exp\left[-{\frac{1}{2}} \left(\alpha'\omega'^2-
2\beta\omega\omega'+\alpha\omega^2 \right) \right]
}
\end{equation}
where
\begin{equation}\displaystyle{
\left.
\begin{array}{lcl}
\alpha &=& {\frac{\sigma_x^2}{c^2}}u_x^2+{\frac{\sigma_y^2}{c^2}}u_y^2+
{\frac{\sigma_z^2}{c^2}}u_z^2+{\frac{\sigma_\tau^2}{c^2}}\\ \\
\alpha' &=& {\frac{\sigma_x^2}{c^2}}u_x'^2+{\frac{\sigma_y^2}{c^2}}u_y'^2+
{\frac{\sigma_z^2}{c^2}}u_z'^2+{\frac{\sigma_\tau^2}{c^2}}\\ \\
{\rm and} \ \ \beta &=& {\frac{\sigma_x^2}{c^2}}u_xu_x'+{\frac{\sigma_y^2}{c^2}}u_yu_y'+
{\frac{\sigma_z^2}{c^2}}u_zu_z'+{\frac{\sigma_\tau^2}{c^2}}
\end{array}
\right\}
}
\end{equation}
Here $\hat{u}=(u_x,u_y,u_z)$ and   $\hat{u'}=(u_x',u_y',u_z')$ are the unit vectors in the 
directions of the incident and scattered fields respectively.

Substituting (1) and (4) in (2), we finally get 
\begin{equation}\displaystyle{ S^{(\infty)}(\omega')=
A'e^{\left[-\frac{1}{2\delta_0'^2}(\omega'-\bar{\omega}_0 )^2\right]} }
\end{equation}
where
\begin{equation}\displaystyle{
\left.
\begin{array}{lcl}
\bar{\omega}_0 &=& \frac{|\beta|\omega_0}{\alpha'+\delta_0^2
(\alpha\alpha'-\beta^2)}\\ \\
\delta_0'^2 &=& \frac{\alpha\delta_0^2+1}{\alpha'+\delta_0^2
(\alpha\alpha'-\beta^2)}\\ \\
 {\rm and} \ \ A' &=& \sqrt{\frac{\pi}{2(\alpha\delta_0^2+1)}}ABA_0\omega_0'^4\delta_0
 exp\left[\frac{|\beta|\omega_0\bar{\omega}_0-\alpha\omega_0^2}
{2(\alpha\delta_0^2+1)}\right]
\end{array}\right\}}
\end{equation}
Though $A'$ depends on $\omega'$, it was approximated by 
James and Wolf ( James and Wolf, 1990)
to be a constant so that $S^{(\infty)}(\omega')$ can be considered to be Gaussian.

The relative frequency shift is defined as 
\begin{equation}\displaystyle{
z=\frac{\omega_0-\bar{\omega}_0}{ \bar{\omega}_0}
}
\end{equation}
where $\omega_0$ and $\bar{\omega}_0$ denote the  unshifted and shifted central frequencies 
respectively. We say that the spectrum is redshifted or blueshifted 
according to whether $~z~>~0~ ~
or~~z~<~0~$ , respectively. Here
\begin{equation}
\displaystyle{
z=\frac{\alpha'+\delta_0^2(\alpha\alpha'-\beta^2)}{|\beta|}-1
}
\end{equation}
\vskip5pt
\noindent
It is important to note that this $z$-number does not depend on the incident frequency, 
$\omega_0$. This is a very important aspect if the mechanism is 
to apply in the 
astronomical domain. Expression (9) implies that the spectrum can be shifted to the blue 
or to the red, according to the sign of the term $\alpha'+\delta_0^2(\alpha\alpha'-\beta^2)~>~
|\beta|~$. To obtain the no-blueshift condition, we use Schwarz's Inequality
which implies that $\alpha\alpha'-\beta^2~\geq~0~$. Thus, we can take 
$$\displaystyle{
 \alpha'~>~|\beta|
 }$$
as the sufficient condition to have only redshift by this mechanism.

Let's now assume that the light in its journey encounters many such scatterers. What we 
observe at the end is the light scattered many times, with an effect as that stated 
above in every individual process. 

 Let there be N scatterers  between the source and the
observer and $z_n$ denote the relative frequency shift after the $n^{th}$ scattering of the 
incident light from the $(n-1)^{th}$ scatterer, with $\omega_n$ and $\omega_{n-1}$ being the 
central frequencies of the incident spectra at  the $n^{th}$ and $(n-1)^{th}$ scatterers. Then by 
definition,
$$\displaystyle{
 z_n =\frac{ \omega_{n-1}-\omega_n}{\omega_n}, ~~~~~~n~=~1,~2,~.~.~.~.,~N
 }$$
or,
$$\displaystyle{
 \frac{\omega_{n-1}}{\omega_n}=1+z_n,~~~~~~n~=~1,~2,~.~.~.~.,~N
 }$$ 
Taking the product over $ n$ from $n ~=~ 1$ to $n~ =~ N$, we get,
$$ \displaystyle{
 \frac{\omega_0}{\omega_N}=(1+z_1)(1+z_2)~.~.~.~.~.~(1+z_N)
 }$$
 The left hand side of the above equation is nothing but the ratio of the
source frequency and the final or  observed frequency $z_f$. Hence,
\begin{equation}\displaystyle{
1 + z_f = (1+z_1)(1+z_2)~.~.~.~.~.~(1+z_N)
}
\end{equation}
\vskip5pt
\noindent
Since the $z$-number due to such effect does not depend upon the central
frequency of the incident spectrum, each $z_i$ depends on $\delta_{i-1}$ only,
not $\omega_{i-1}$ [here $\omega_j$ and $\delta_j$ denote the central frequency
and the width of the incident spectrum at $(j+1)^{th }$ scatterer]. To find the
exact dependence we first calculate the broadening of the spectrum after N number of scatterings.

Here, we are considering the multiple scattering on the assumption that the
scatterers are mutually incoherent. In this case the cross
terms in the spectrum of the scattered field are zero. In other words we are not
including higher order scatterings. Moreover, we are also
 considering small angle scatterings so that there will be small deviation from
the actual path of  light. As a result the superimposed  spectra from different
scatterings will result in a signal in the forward direction only. This is similar to
the single scattering  result of James and Wolf.

\vskip8pt
\subsection{Effect of Multiple Scatterings on the Spectral Line Width}
\vskip5pt
From the second equation in (7), we can easily write,
\begin{equation}\displaystyle{
\left.
\begin{array}{lcl}
\delta_{n+1}^2 & = &
\frac{\alpha\delta_n^2+1}{\alpha'+(\alpha\alpha'-\beta^2)\delta_n^2} \\ \\
& = & \left(\frac{\alpha\delta_n^2+1}{\alpha'}\right)\left[1+\delta_n^2\left(
\frac{\alpha\alpha'-\beta^2}{\alpha'}\right)\right]^{-1} 
\end{array}
\right\}}
\end{equation}

From (13), we can also write
\begin{equation}
\displaystyle{
\omega_{n+1} = 
\frac{\omega_n|\beta|}{\alpha'+(\alpha\alpha'-\beta^2)\delta_n^2} 
}
\end{equation}

Then from (11) \& (12), we can write
\begin{equation}
\displaystyle{
\left.
\begin{array}{rcl}
z_{n+1}&=&\frac{\omega_n-\omega_{n+1}}{\omega_{n+1}}\\ \\
&=&\frac{\alpha'+(\alpha\alpha'-\beta^2)\delta_n^2}{|\beta|}-1\\ \\
&=&\frac{\alpha'}{|\beta|}\left\{1+\left(\frac{\alpha\alpha'-\beta^2}{\alpha'}
\right)\delta_n^2\right\}-1
\end{array}
\right\}}
\end{equation}
\noindent
Let's assume that the redshift per scattering process is very small, {\it i.e.,}
$$0~<~\epsilon ~ = ~ z_{n+1}~~<<~1 $$ for all $n$.
\vskip5pt
\noindent
Then,
$$\displaystyle{
\begin{array}{lrcl}
& 1+\epsilon & = & \frac{\alpha'}{|\beta|}\left\{1+\left(\frac{\alpha\alpha'
-\beta^2}{\alpha'}\right)\delta_n^2\right\}\\ \\
{\rm or,}& (1+\epsilon)\frac{|\beta|}{\alpha'} & = & 1+\left(\frac{\alpha\alpha'-
\beta^2}{\alpha'}\right)\delta_n^2
\end{array}
}$$
In order to satisfy this condition and in order to have a redshift
 only ( or positive z),
we see that the first factor $\frac{\alpha'}{|\beta|}$ in the right term cannot be much 
larger than 1, and, more important, 
\begin{equation}
 \left(\frac{\alpha\alpha'-\beta^2}{\alpha'}\right)\delta_n^2~~<<~~1
\end{equation}

In that case, from (11), after neglecting higher order terms, the expression for 
$~~\delta_{n+1}^2~~$ can be well approximated as:
$$\displaystyle{
\delta_{n+1}^2  \approx
 \left(\frac{\alpha\delta_n^2+1}{\alpha'}\right)\left[1-\delta_n^2\left(
\frac{\alpha\alpha'-\beta^2}{\alpha'}\right)\right]
}$$
which, after carrying out a  simplification, gives a very important recurrence relation:
\begin{equation}\displaystyle{
\delta_{n+1}^2 =  \frac{1}{\alpha'}+\frac{\beta^2}{\alpha'^2}\delta_n^2.
}
\end{equation}

Therefore,
$$\displaystyle{
\begin{array}{lcl}
\delta_{n+1}^2 & = & \frac{1}{\alpha'}+\frac{\beta^2}{\alpha'^2}\delta_n^2 \\
\\ 
& = & \frac{1}{\alpha'}+\frac{\beta^2}{\alpha'^2}[ \frac{1}{\alpha'}
+\frac{\beta^2}{\alpha'^2}\delta_{n-1}^2] \\ \\
& = & \left(\frac{\beta^2}{\alpha'^2}\right)^2\delta_{n-1}^2
+\frac{1}{\alpha'}\left(1+\frac{\beta^2}{\alpha'^2}\right) \\ \\
& .~. & .~.~.~.~.~.~.~.~.~.~.~.~.~.~  \\ \\
& = & \left(\frac{\beta^2}{\alpha'^2}\right)^{n+1}\delta_0^2
+\frac{1}{\alpha'}\left(1+\frac{\beta^2}{\alpha'^2}+~.~.~.~.~
\frac{\beta^{2n}}{\alpha'^{2n}}\right).
\end{array}
}$$

Thus 
\begin{equation}\displaystyle{
\delta_{N+1}^2  =
\left(\frac{\beta^2}{\alpha'^2}\right)^{N+1}\delta_0^2
+\frac{1}{\alpha'}\left(1+\frac{\beta^2}{\alpha'^2}+~.~.~.~.~
\frac{\beta^{2N}}{\alpha'^{2N}}\right).
}
\end{equation}

As the number of scattering increases, the width  of the spectrum obviously
increases and the most important topic to be considered is whether this width
is below some tolerance limit or not, from the observational point of view.
There may be several measures of this tolerance limit. One of them is the {\it
Sharpness Ratio}, defined as
$$ Q=\frac{\omega_f}{\delta_f}$$
where $~\omega_f ~$ \& $~ \delta_f ~$ are the mean frequency \& the width of
the observed spectrum.

After $N$ number of scatterings, this sharpness ratio, say $Q_N$, is given by
the following recurrence relation :
$$Q_{N+1}=Q_N \sqrt{\frac{\alpha'}{\alpha'+(\alpha\alpha'-\beta^2)\delta_N^2}-
\frac{1}{\alpha\delta_N^2+1}}.$$ 
It is easy to verify that the expression under the square root lies between 0 \&
1. Therefore, $Q_{N+1}~~<~~Q_N$, and the line is broadened as the scattering 
proceedss  on . 

Under the sufficient
condition of redshift [i.e., $|\bt|~<~\al'$ ](Datta et al.,1998c) it
was shown that in the observed spectrum
\begin{equation}
\Dl\om_{n+1} \gg \dl_n
\end{equation}
\noindent
if the following condition holds:
\begin{equation}
\frac{\dl_n\om_0(\al\al'-\bt^2)}{\al'+(\al\al'-\bt^2)\dl_n^2} \gg 1
\end{equation}
where $\om_0$ is the source frequency.

The relation (18) signifies that the shift is more prominent than the effective
broadening so that the  spectral lines are observable and can  be analyzed. If, 
on the other hand ,the
broadening is higher than the shift of the spectral line, it will be impossible
to detect the shift from the blurred spectrum. Hence we can take  relation
(18) to be one of the conditions necessary for the observed spectrum to be
analyzable. For large $N$ (i.e., $N \rightarrow \infty $ ), the series in the second 
term of right hand side of (16) converges to a finite sum and we get

$$ \displaystyle{
\delta_{N+1}^2  =
\left(\frac{\beta^2}{\alpha'^2}\right)^{N+1}\delta_0^2
+\frac{\alpha'}{\alpha'^2 - \beta^2}.}$$

If $\delta_0$ is considered as arising out of Doppler broadening only, we can 
estimate  $\delta_{\rm Dop}  \sim 10^9$  for $T = 10^4$ K. On the other hand, for an anisotropic 
medium, we can take $\sigma_x = \sigma_y = 3.42 \times 10^{-1}, \ \sigma_z = 
8.73 \times 10^{-1}, \ \alpha' = 8.68 \times 10^{-30}, \ \alpha = 8.536 \times
 10^{-30} {\rm and} \beta = 8.607 \times 10^{-30}$  
for $\theta = 15^0$(James and Wolf,1990). Then the 
second term  of the above expression will be much larger than the first 
term, and effectively, Doppler broadening can be neglected in comparison to 
that due to multiple scattering effect.
\vskip5pt
\noindent
Now if we consider the other condition, $ i.e.,~~\al'~<~|\beta|$, the series in
(16) will be a divergent one and $\dl_{N+1}^2$ will be finitely large for large
but finite $N$. However, if the condition (17) is to be satisfied, then the
shift in 
frequency will be larger than the width of the spectral lines. In that case
 the
condition $\al'~<~|\beta|$ indicates that blueshift may also be observed but
the width of the spectral lines can be large enough depending on how large the
number
of collisions is. So in general, the blueshifted lines should be of larger widths
than the  redshifted lines and may not be as easily observable.

\vskip10pt
\section{\bf Critical Source Frequency}
\vskip5pt
\noindent
Rearranging  equation (18) we get,
\beq
\left(\dl_n-\frac{\om_0}{2}\right)^2 \ll
\frac{\om_0^2}{4}-\frac{\al'}{\al\al'-\bt^2} 
\eeq
Since the left side is non-negative, the right side must be positive. Moreover,
since the mean frequency of any source is always positive, {\it i.e.,}
$\om_0 \geq 0$, we must have
\beq
\om_0 \gg \sqrt{\frac{4\al'}{\al\al'-\bt^2}}.
\eeq
We take the right side of this inequality to be the {\bf \it critical source
frequency } ${\bf \om_c}$ which is defined here as.
\beq
\om_c = \sqrt{\frac{4\al'}{\al\al'-\bt^2}}.
\eeq
Thus for a particular medium between the source and the observer, the  critical
 source frequency is the lower limit of the frequency of any
source whose spectrum can be clearly analyzed. In other words, the shift of any
spectral line from a source with frequency less than the critical source frequency for that
particular medium cannot be detected due to its high broadening.
\vskip5pt
\noindent
 We can  now classify the spectra of the different sources,  from which light 
comes
to us after passing through a scattering medium characterized by the parameters 
$\al$, $\al'$,
$\bt$. If we allow only small angle scattering in order to get prominent
spectra, according to  the Wolf mechanism,  they will  either be blueshifted or
redshifted. The redshift of spectral lines may or may not be detected according 
to whether
the condition (20) does or does not hold. In this way those sources whose spectra
are redshifted, are classified in two cases, {\it viz.,}  ${\bf \om_0~~>~~\om_c}$  and
{\bf $\om_0~~\leq~~\om_c$}. In the first case, the shifts of the spectral lines can
be easily detected due to condition (17). But in the later case, the spectra
will suffer from the  resultant blurring.
\vskip10pt
\subsection{\bf Critical Source Frequency and Anisotropy}
\vskip5pt
\noindent
The anisotropy discussed in this paper is statistical in nature. We have considered
both temporal as well as spatial correlations to induce such spectral changes.
The spatial anisotropy has been characterized by unequal correlation lengths along three
mutually perpendicular directions. We have used the term {\it strong anisotropy}
 in  a  particular direction ( among the above three) to specify that the correlation
length along that direction is very large compared to that of the other two directions.
We now find the critical source frequency in both the isotropic and anisotropic cases.

{\it Spatially Isotropic Case}

If the spatial correlation lengths are equal in every direction , i.e. ${ \sigma_x 
= \sigma_y = \sigma_z =\sigma }$ then the parameters ${\alpha, \alpha\prime , \beta}$
reduce to 
\beq
\alpha = \alpha^\prime = \frac{{\sigma}^2}{c^2} + \frac{{\sigma_\tau}^2}{c^2}
 ; \ \ \ \beta = \frac{{\sigma}^2}{c^2} \cos\theta +
 \frac{{\sigma_\tau}^2}{c^2}
\eeq
Therefore from (21) we get
\beq
\omega_c = \frac{2c}{\sigma} \sqrt {\frac{\sigma^2 + {\sigma_\tau}^2}
{{(1 - \cos\theta)}{{\sigma^2(1 + \cos\theta)+ 2 {\sigma_\tau}^2}}}}
\eeq

Therefore,
$$\displaystyle{
\begin{array}{lcl}
 \omega_c & = & \frac{2c}{\sigma} cosec \theta ; \ \ {\rm if} \ \ \sigma >>
 \sigma_\tau \nonumber \\
 \omega_c & = & \frac{2c}{\sigma}\sqrt{\frac{1}{2(1-\cos\theta)}} \ \ 
{\rm if}\ \sigma << \sigma_\tau 
\end{array}}
$$

In both  cases $\sigma_c$ varies inversely with $\sigma$. This is
equivalent to taking $ \sigma_\tau = 0$ i.e. for white noise.

{\it Spatially Anisotropic Case }

Let there be a strong anisotropy along, say, the x-axis. Then
$\sigma_x >> \sigma_y , \sigma_z$. Consequently,
\beq
\alpha = \frac{\sigma^2}{c^2} + \frac{\sigma_{\tau}^2}{c^2}; \ \
\alpha^\prime = \frac{\sigma^2}{c^2} {u_x^\prime}^2 + \frac{{\sigma_\tau}
^2}{c^2}; \ \
\beta = \frac{\sigma^2}{c^2} u_x {u_x^\prime} + \frac{{\sigma_\tau}^2}{c^2}
\eeq
 We can define 
 $$k = \frac{\sigma_x}{\sigma_\tau}$$ as a ratio of the correlation lengths along the spatial and temporal
directions. We now consider the variation of $\omega_c$ with $k$ in the following two case :

1. $\sigma_\tau$ is fixed ; $k$ varies with $\sigma_x$
 
2. $\sigma_x$ is fixed ; $k$ varies inversely with $\sigma_\tau$.\\
The  nature of the variation of $\omega_c$ against $k$ in the above two
cases has been studied by Datta et al.(Datta et al.1998a,b).
We can easily find the critical source frequency for
$ k = 1$ as
\beq
\omega_c = \frac{2c \sqrt{{u_x ^\prime}^2 + 1}}{\sigma
{|u_x - u_x^\prime |}}
\eeq
From (24) and (25) we can say that the stronger the anisotropy, the lower the
critical source frequency and hence the greater the scope of more and more spectra
to be anlyzable. This induces a special type of screening that we are now going to
discuss.

\section{\bf Screening Effect Induced by  the Critical Source Frequency}

Collisional mechanism of light beams give some sort of screening which may
or may not show frequency shift of the spectral lines. Here the scatterer is the
medium with its dielectric susceptibility randomly fluctuating both spatially
and temporally.  The critical source frequency induces a special type of screening
which not only changes the colour of the incident lines but also broadens their
widths. Specifically speaking, when an incident line with its peak frequency
$\omega_0$ encounters the scattering medium whose critical source frequency
is $\omega_c$ , then
\begin{enumerate}
\item  frequency shift occurs 
\item   the spectral width increases and this increased width will be less than or
greater than the frequency shift according as $\omega_0$ is greater or less than
$\omega_c$.
\end{enumerate}
Let a spectral line with wavelength greater than the critical wavelength of the
medium propagate and hence be scattered. The scattered line, according to the screening effect, will be of width greater than
the shift of that line. Therefore, the observer will find the broadening of the 
line more prominent than its shift and hence it will be difficult to analyze the line.
On the other hand, if the wavelength of the line is less than the critical 
wavelength, its shift will dominate the broadening.
In the case of strong spatial anisotropy ( without loss of generality, we may assume
it along the x-axis) it can be shown that the maximum value of this critical wavelength for a 
particular medium ( i.e. for a particular set of the spatial correlation
lengths) is given by the following relation:

\beq
\lambda_c = \pi \sqrt{\sigma_x^2 + \sigma_{\tau}^2}
\eeq
Thus this type of screening involves only two parameters - the spatial and temporal
correlation lengths. If we take a particular medium and may know its spatial
and temporal correlation lengths , we can easily find that particular wavelength
from (26) which enables us to get the nature of the screening. To be more specific, we can find
out the wavelength zone beyond which the anlyzability of the spectral lines will
be difficult.
\vskip10pt
\noindent

\section{\bf Possible Implications for Quasar Redshifts}
\vskip5pt
The screening effect as discussed above might play a significant
role in observations of quasars  as well as in  the laboratory.
If we consider the spectra originating from distant quasars,
it might be possible to explain the large redshift and broadening
by this type of screening. We may try to visualize
this at least in a simple fashion for the redshifts in the case of 
NGC 4319 Galaxy and Markarian 205 Quasar pair as follows.

\vskip5pt
\noindent
\subsection{\it Screening Effect and  Galaxy-Quasar Pair}
\vskip5pt
\noindent

The above mentioned screening effect gives in this case the simple explanation
for the redshift controversy in  a galaxy-quasar pair ( Sulentic, 1989) 
[  NGC 4319  is the Galaxy and 
 Markarian 205 the quasar].
They appear to be connected in very deep photographs but the quasar's redshift
is found to be much higher ( $z$-number is almost 12 times ) than that of the
galaxy. This empirical evidence is normally taken as one of the strongest
arguments in favour of non-cosmological redshifts. Weedman ( Weedman,1986 )
made a critical survey of this situation using statistical arguments for
comparisons of positions in the sky for quasars and galaxies. It has been
claimed ( Weedman, 1986) that if one uses this kind of statistical test
for several samples, none  shows evidence for association of high
redshift quasars with low redshift galaxies. This impacts on hypotheses of both
non-cosmological redshifts and gravitational lensing. Here, we want to
emphasize that this kind of association of high redshift quasars with low
redshift galaxies is a possibility in our theory of shift of the spectral lines
by multiple scatterings. We can try to visualise the situation in the following
way :

We know very little little about the atmospheres surrounding those distant galaxies
and quasars. We nevertheless, assume that they are very different. Therefore the medium
through which we observe the quasar is  much different from that for the
galaxy.  Two sets of parameters $(~~\al,~~~\al',~~~\bt~~)$ characterising the
media thus give rise to two different critical source frequencies, say $\om_G$
( for galaxy ) and $\om_Q$ ( for quasar ). Now if the lines emitted from the
galaxy have frequencies less than $\om_G$, then the Wolf-shift cannot be
observed. On the other hand, if those lines emitted from the quasar have
greater frequencies than $\om_Q$, prominent Wolf-shifts will be observed. The
Doppler shift ( or the shift as interpreted in the expanding Universe picture ) might be
present in both cases. Thus the net shift observed in the case of the quasar
will be higher than  in the case of the galaxy. Let the $z$-number of the
galaxy be ${(\rm say z_1=)} a$ and the $z$-number of the quasar due to the Wolf effect be ${(\rm say z_2=)} b$. Again the resultant $z$ number can be written
as $ 1 + z = (1+Z_1)(1+z_2)$. Then

$$ \begin{array}{lrcl} &12a&=&a+b+ab\\
\Rightarrow&b&=&\frac{11a}{1+a}
\end{array}$$
This indicates that the distance of the quasar is just the same as that of the
galaxy, but the shift in the first case is much higher than that of the later.
We can envisage this phenomena as some sort of {\it screening} operation for the
Wolf Effect. In other words, new type of screening arises due to the nature of
the medium . This will  be studied in details in future work.
Lastly, we should mention that the critical source frequency relating 
to the screening effect which plays a crucial role in explaining both  redhsift and
spectral width can be tested in {\it laboratory experiments}. 

\vskip20pt
\noindent
{\bf Acknowledgements:} One of the authors ( S.D.) greatly thanks World
Laboratory, Laussane for financial support during this work and Prof. B. K.
Datta ( Director, World Laboratory, Calcutta Branch ) for encouragement. 
The author(S.R.) is indebted to Prof. Jack Sulentic,University of Alabama,
for valuable suggestions and comments. The authors like to thank the referee
for valuable suggestions.

\newpage

{\bf REFERENCES}

\begin{enumerate}

\item  Born, M.\& Wolf,E. 1997, Principle of Optics, 6th edition, 
       Pergamon, Oxford.
\item  Datta S, Roy S, Roy M \& Moles M, 1998a,, Int. Jour. of Theo. Phys., 
       {\bf 37},N4, 1313.  
\item  Datta S, Roy S, Roy M \& Moles M, 1998b, Int.Jour.Theort.Phys., 
      {\bf 37},N5, 1469. 
\item  Datta S, Roy S, Roy M and Moles M, 1998c, Phys.Rev.A,{\bf 58},720

\item  James D.F.V and Wolf E, 1990, Phys. Lett. A, {\bf 146}, 167.

\item  Roy S., Kafatos M. and Datta S. 1999, Broadening
       of Spectral Lines Due to Dynamic Multiple Scattering and The 
       Tully-Fisher Realation , Physical Review A ,{\bf 60 },273 .
\item  Sulentic J.W.1989, ApJ.{\bf 345}, 54.
\item  Weedman D.W.,1986, Quasar Astronomy,  Cambridge University
       Press.
\item  Wolf Emil ,1998, The Redshift Controversy and Correlation-Induced                     Spectral Changes" Eds.Pratesi R. and Ronchi L.
                       Spectral changes, Eds.Pratesi R. and Ronchi I.
                      in  Waves, Information and Foundations of Physics
                     {\bf 60}, p.41 .
\item  Wolf E.,1986,Phys.Rev.Lett.{\bf 56}, 1370.
\item  Wolf E. and James D.F.V.,1996 ,Rep. Progr. Phys.{\bf 59}, 771.

\end{enumerate}

\end{document}